\documentclass[12pt]{article}
\usepackage{amssymb}

\usepackage{cite} 
\usepackage[english]{babel}
\usepackage{hyperref}
\usepackage{xcolor}
\usepackage{bbold} 
\renewcommand{\theequation}{\arabic{equation}}
\newcommand{\be}{\begin{equation}}
\newcommand{\ee}{\end{equation}}
\newcommand{\bea}{\begin{array}}
\newcommand{\ea}{\end{array}}
\newcommand{\beqa}{\begin{eqnarray}}
\newcommand{\eeqa}{\end{eqnarray}}
\newcommand{\bean}{\begin{eqnarray*}}
\newcommand{\eean}{\end{eqnarray*}}

\def\up#1{\leavevmode \raise.16ex\hbox{#1}}

\newcommand{\gapproxeq}{\lower
.7ex\hbox{$\;\stackrel{\textstyle >}{\sim}\;$}}
\newcommand{\lapproxeq}{\lower .7ex\hbox{$\;\stackrel
{\textstyle <}{\sim}\;$}}
\renewcommand{\theequation}{\thesection.\arabic{equation}}
\newcounter{appendice}
\newcommand{\appendice}
{
\setcounter{equation}{0}
\renewcommand{\theequation}{\Alph{appendice}.\arabic{equation}}
\addtocounter{appendice}{1}
{\Large{\bf Appendix \Alph{appendice}}}
}
\def\thebibliography#1{{\bf REFERENCES\markboth
{REFERENCES}{REFERENCES}}\list
{[\arabic{enumi}]}{\settowidth\labelwidth{[#1]}\leftmargin\labelwidth
\advance\leftmargin\labelsep
\usecounter{enumi}}
\def\newblock{\hskip .11em plus .33em minus -.07em}
\sloppy
\sfcode`\.=1000\relax}

\def\BI{{\rm 1\!l}}

\newcommand{\al}[1]{\textcolor{black}{#1}}

\begin{document}

\centerline{\LARGE Embedding Space Approach to JT Gravity}
\vskip 1cm

\centerline{ A. Pinzul${}^1$\footnote{apinzul@unb.br}, A. Stern$^{2}$\footnote{astern@ua.edu} and   Chuang Xu$^{2}\footnote{cxu24@crimson.ua.edu} $ }

\vskip 1.5 cm

\begin{center}
  {${}^1$ Universidade  de Bras\'{\i}lia, Instituto de F\'{\i}sica\\
70910-900, Bras\'{\i}lia, DF, Brasil\\
and\\
International Center of Physics\\
C.P. 04667, Bras\'{\i}lia, DF, Brazil
\\}
\end{center}
\begin{center}
  {${}^2$ Department of Physics, University of Alabama,\\ Tuscaloosa,
Alabama 35487, USA\\}
\end{center}

\vskip 0.5cm

\abstract{We present a coordinate-free background space construction of Euclidean Jackiw-Teitelboim  gravity. It is written as a gauge theory obtained from  the Killing vectors and conformal Killing vectors of a hyperboloid embedded in a three dimensional background. A novel feature of the gauge theory  is that   vanishing  field strength does not necessarily imply that the gauge potentials  \al{(written in the embedding space)} are pure gauges, not even locally. \al{On the other hand, the projection of the potentials along the hyperboloid are pure gauges, as is the case in the standard approach.}  As is usual,  metric tensors are dynamically generated from the classical solutions of the theory, which here do not rely on coordinate charts on the two-dimensional surface.  We find a special class of  solutions whereby the derived metric tensor on the surface is the induced metric from the background space.  The gauge theory construction given here has a  natural generalization  to a non-commutative space, which does not require the use of coordinates, symbols, or a star product. }

\section{Introduction}
Jackiw-Teitelboim (JT) gravity\cite{T1983,J1985} is a theory of two-dimensional gravity with nonzero constant local curvature.  In the original formulation of the theory, after assuming some coordinate charts on the two-dimensional manifold, one introduces dynamics in terms of the metric tensor and a space-time scalar. JT gravity is also commonly written as an $SU(1,1)$ gauge theory, with the field strength related to the torsion and curvature.\cite{FUKUYAMA1985259,Isler,CHAMSEDDINE198975,Jackiw:1992bw,Celada:2016jdt}
 In recent years research in JT gravity has shown connections to duality \cite{AP2015,J2016,MSY2016} and random matrix theory\cite{SSS2019,SW}.  
There has also been interest in alternative formulations of JT gravity,  c.f.\cite{MST2021,LU2023}.   Of course, an ambitious goal for an alternative formulation is  to  have a  theory of gravity that can be suitably extended up to the Planck scale. Since there is a  general consensus that the usual notions of space-time break down as one approaches the Planck scale, it could be useful to start with a coordinate-free description of the theory. With this in mind we show that JT gravity can be formulated without the use  of coordinate charts on the space-time manifold.

The basic idea is quite simple:
In its original formulation, the nonvanishing constant curvature in JT gravity is a result of  classical dynamics. From the $SU(1,1)$ gauge theory perspective, the field strength vanishes as a result of the equations of motion, implying a torsionless space of constant curvature.  Alternatively, one can  introduce a two-dimensional surface of constant curvature, i.e., a hyperboloid, \textit{ab initio}, and embed it in some higher dimensional background space where the isometry transformations are more transparent. A  gauge theory description of JT gravity can once again be introduced, now however, with the gauge fields written as functions of only the embedding into the background, rather than coordinates  on the two-dimensional  subsurface.

The gauge theory presented here  is constructed  starting from the background, which we choose to be  three-dimensional Minkowski space.  It  has some novel features.  It  relies on the use of the Killing vectors, as well as conformal Killing vectors, on the embedded hyperboloid.  The Killing vectors  preserve the induced metric tensor and generate an isometry group ${\cal I}$ which is isomorphic to $SO(2,1)$. Their action can be expressed in terms of the natural Poisson structure on the hyperboloid. The Killing vectors, along with  the conformal Killing vectors,  span the $so(2,2)$ Lie-algebra.  The relevant gauge group is, once again, $SU(1,1)$.
We shall give two different formulations of the $SU(1,1)$ gauge theory, labeled I and II.  The Killing vectors and conformal Killing vectors play opposite roles in the two formulations.  In    I,  gauge transformations are generated using the Killing vectors, and the field strength is defined in terms of the conformal Killing vectors.  In II, gauge transformations are generated using the conformal Killing vectors, and the field strength is defined in terms of the Killing vectors.  While the two formulations are dynamically equivalent,  they have different features.  Formulation II has the advantage that a straightforward projection to the subsurface gives the standard gauge theory description of JT gravity.   Formulation I, on the other hand, has a natural extension to the non-commutative hyperboloid.  \al{We note that for both formulations I and II, the space of gauge connections is larger than in the standard formulation because they are defined in the three-dimensional embedding space, rather than only along the two-dimensional submanifold.  We shall refer to the gauge potentials in formulations I and II as embedding space potentials.}

Another novel feature of the gauge theory presented here is that the vanishing  of the field strength does not necessarily imply that the \al{ embedding space}  potentials  are pure gauges, not even for a local region on the hyperboloid.   In general, the solution to the zero field strength condition  is the sum of a pure gauge \al{directed  along the hyperboloid}  and a component of the potential which is normal to the hyperboloid. The normal component projects to the space-time scalar field of the standard formulation. {\it  So here  the   scalar field  is a result of the  construction in the higher dimensional background space, and there  is no need to introduce it by hand.} 

Zweibein and spin connection one forms can be identified, as usual, from the pure gauge components of the potentials, expressed in terms of  some $SU(1,1)$-valued function $V$, which are then used to define the  metric tensor for the theory.
Provided that it is  nonsingular, this metric tensor $g_{\mu\nu}(V)$ describes two-dimensional de Sitter $dS_2$ or anti-de Sitter space $AdS_2$. (We choose the latter in this article. More specifically, we  specialize to  Euclidean $AdS_2$, although there are no formal obstacles to repeating the analysis  presented here for the Lorentzian analogue.)  However,  $g_{\mu\nu}(V)$ need not be identical to the induced metric, even though it is related to the latter (at least, locally) via a diffeomorphism.  Moreover, the Killing vectors and conformal Killing vectors associated with $g_{\mu\nu}(V)$, do not, in general, agree with those of the induced metric.   The different choices for the $SU(1,1)$-valued function $V$ lead, in general, to different metrics $g_{\mu\nu}(V)$.
They can be  categorized in terms of how $V$  transforms under the simultaneous action of the isometry group ${\cal I}$ and a global $SU(1,1)$ group $G$, i.e., under the action of the diagonal subgroup $({\cal I} \otimes G)_{diag}$ of the product  group ${\cal I} \otimes G$. One might perhaps expect the induced metric tensor to be associated with those group elements $V$ with the highest degree of symmetry under $({\cal I} \otimes G)_{diag}$, but this turns out not to be the case.
We find instead that the induced metric tensor agrees with  $g_{\mu\nu}(V)$  for those  $V$ that are invariant under only the $U(1)$ subgroup of $({\cal I} \otimes G)_{diag}$.

The outline for the rest of this article is the following.  In section 2, we first introduce the embedding, and express   the action of the Killing vectors and conformal Killing vectors  in terms of the natural Poisson structure. We then give the two equivalent coordinate-free descriptions of  $SU(1,1)$ gauge theory on the embedded surface, and show how to recover JT gravity. In section 3, we examine two  types of pure gauge solutions and their $g_{\mu\nu}(V)$. For the first type, $V$ is invariant under $({\cal I} \otimes G)_{diag}$, while for the second type, $V$ is  invariant under only the $U(1)$ subgroup of $({\cal I} \otimes G)_{diag}$.  As stated above, the induced metric tensor is contained in the latter.
In section 4 we briefly outline the  natural generalization of the gauge theory to a noncommutative space.  The extension necessitates  enlarging the $SU(1,1)$ gauge group to $U(1,1)$, but it does not require the introduction of coordinates, symbols or a star product.
Some concluding remarks are offered in section 5.  In the appendix we show that the Killing vectors defined in section 2 annihilate the induced metric tensor, and we give the action of the conformal Killing vectors on the induced metric.

\section{Embedding space formulation of JT gravity}
\setcounter{equation}{0}

\subsection{Euclidean $AdS_2$}
In this article we specialize to the Euclidean version of JT gravity. It is written on Euclidean $AdS_2$. The latter can be defined by  restricting to a single sheeted hyperboloid embedded in $3$D Minkowski space. Denoting quasi-Cartesian coordinates of $3$D Minkowski space by $X^a$, $a=0,1,2$, a single sheeted hyperboloid is given by
\be
X^a X_a=-1 \;,\qquad X^0>1\label{EXaXaismnsn}\;.
\ee
Here indices $a,b,c,..$ are raised and lowered using the flat metric
\be
\eta={\rm diag}(-++)\;.\label{Ebkgrnd}
\ee
The induced metric on the surface is
\be
g^{\mbox{\tiny ind}}_{\mu\nu}=\partial_\mu X^a\partial_\nu X_a\;,\label{indmtric}
\ee
where $\partial_\mu$   denote partial  derivatives  with respect to coordinates on the surface and $\mu ,\nu =0,1$.
The result that it has a Euclidean signature follows from the inequality
\be
\det  g^{\mbox{\tiny ind}}_{\mu\nu}=\frac 1{(X^0)^2}\left(\epsilon^{\mu\nu}\partial_\mu X^1\partial_\nu X^2\right)^2\;\ge\;0 \;,
\ee
where $\epsilon^{\mu\nu}=\epsilon_{\mu\nu}$ is totally antisymmetric, with $\epsilon^{01}=1$.

The isometry group for $AdS_2$ is $SO(2,1)$. It is generated by Killing vectors $K^a$, $a=0,1,2$, satisfying
\be
[K^a,K^b]=\epsilon^{abc} K_c\;,\label{so21Kil}
\ee
where $\epsilon^{abc}$ is totally antisymmetric, with $\epsilon^{012}=-\epsilon_{012}=1$. The defining relation (\ref{EXaXaismnsn}), and the induced metric (\ref{indmtric}) are preserved upon assuming the following action of $K^a$ on the embedding coordinates\footnote{The parenthesis in the expression below indicate that  the Killing vectors  do not act to the right of "$)$".}
\be
(K^aX^b)=\epsilon^{abc} X_c \;.\label{KaonXb}
\ee

A natural Poisson structure can be written on the hyperboloids which is preserved by the action of the Killing vectors, namely
\be
\{X^a,X^b\}=\epsilon^{abc} X_c\;.\label{su11pba}
\ee
It follows from (\ref{EXaXaismnsn}) that
\be
\{ X_a,{\cal G}\} X^a=0\;,\label{XFXidnt}
\ee
for any function ${\cal G}$ of the embedding coordinates. Moreover, from (\ref{KaonXb}) we can implement the action of the Killing vectors using the Poisson bracket
\be
(K^a{\cal G})=\{ X^a,{\cal G}\}\label{onedfKv}\;,
\ee
for any  function ${\cal G}(X)$. The commutator (\ref{so21Kil}) then follows from the Jacobi identity for the Poisson bracket. In the appendix  we show that $K^a$ satisfies the Killing equations for the induced metric (\ref{indmtric}), i.e.,
\be
{\cal L}_{K^a}\,g^{\mbox{\tiny ind}}_{\mu\nu}=0\;,\label{Lieong}
\ee
with ${\cal L}$ denoting the Lie derivative.

In addition to $K^a$, one can define {\it conformal} Killing vectors
\be
W_a=\epsilon_{abc}X^b K^c\label{conKillers}\;.
\ee
Both $K_a$ and $W^a$ play a role in defining the gauge theory associated with JT gravity.
When acting on the embedding coordinates spanning the hyperboloid it gives $(W_a X^b)=X_a X^b+\delta_a^b\;$. In the appendix  we show that the action of  the Lie derivative of $W_a$ on the induced metric tensor is
\be
{\cal L}_{W_a}\,g^{\mbox{\tiny ind}}_{\mu\nu}=2X_a\, g^{\mbox{\tiny ind}}_{\mu\nu}\;.\label{rsclmetrc}
\ee
$W_a$ satisfy the commutation relations
\be
[K_a,W_b]=\epsilon_{abc} W^c\;,\quad  [W_a,W_b]=-\epsilon_{abc} K^c \;.\label{othrctrs}
\ee
Then $W_a$, together with $K_a$, give a basis for  $so(2,2)\simeq su(1,1) \otimes su(1,1)$. More concretely, the two copies of $su(1,1)$ are generated by the mutually commuting sets of generators, $\frac 12\left(K_a + i W_a\right)$ and $\frac 12\left(K_a - i W_a\right)$.

\subsection{$SU(1,1)$ gauge theory}\label{SU(1,1)GT}

In this section we present two equivalent formulations of $SU(1,1)$ gauge theory, \al{ labeled I and II,   written in the  three-dimensional embedding space,} each having its own particular advantages.  The  embedding space gauge potentials defining these theories are \al{thus three-dimensional space-time vectors, which will be denoted by ${\cal A}$ in I, and $A$ in II.  They differ in that} gauge transformations of ${\cal A}$ are implemented using the  Killing vectors, while gauge transformations of $A$ are implemented using the conformal  Killing vectors.  \al{On the other hand, the field strength for the former is constructed  from conformal  Killing vectors, while in the latter it is constructed  from  Killing vectors.}

\subsubsection{Formulation I} \label{FormI}

{Next let us introduce  $su(1,1)-$valued  \al{embedding space} potentials $\mathcal{A}=\{\mathcal{A}_a, \;a=0,1,2\}$, which are defined to transform as vectors under the action of the $SO(2,1)$ isometry group. First, we will proceed in the standard heuristic way. Namely, consider a ``matter'' field, $\Phi$, in, say, fundamental representation of $SU(1,1)$. } The natural derivatives in the flat $3D$ background are translation generators.  Here, we are instead interested in derivatives that are compatible with the isometries of $AdS_2$, which  are {given by the Killing vectors in (\ref{onedfKv}), i.e., $(K^a \Phi) = \{X^a , \Phi\}$. The local action of $SU(1,1)$ on the field is not covariant
\be
K^a U \Phi \neq U K^a \Phi\ ,\ \mathrm{for\ any\ local}\ U\in SU(1,1)\;.
\ee
Introduce $\mathcal{A}=\{\mathcal{A}_a, \;a=0,1,2\}$ by requiring
\be
( K_a + \mathcal{A}_a ) U \Phi = U (K^a  + \mathcal{A}'_a ) \Phi\ ,\ \mathrm{for\ any\ local}\ U\in SU(1,1)\;.
\ee
We immediately see that $\mathcal{A}_a$ transforms as the usual gauge field
\be\label{Gauge}
\mathcal{A}'_a = U^{-1} K_a U + U^{-1}\mathcal{A}_a U\;.
\ee
The fact that the first term is written in terms of Killing vectors, and not simply partial derivatives will be crucial below.  From (\ref{Gauge}) and the identity $X^b K_b =0$, we see that the normal component of the potentials, $X^b \mathcal{A}_b $, transforms in the adjoint representation, i.e. $X^b \mathcal{A}'_b = U^{-1}X^b \mathcal{A}_b U$. Continuing, formally, with constructing the field strength, we have
\be\label{Field}
\mathcal{F}_{ab}\Phi := [K_a +\mathcal{A}_a ,K_b +\mathcal{A}_b]\Phi = \left( \epsilon_{abc}K^c + (K_a  \mathcal{A}_b)- ( K_b  \mathcal{A}_a) +[\mathcal{A}_a , \mathcal{A}_b] \right)\Phi\;.
\ee
We see that due to the Lie-algebra type relations (\ref{so21Kil}), our would-be field strength} is a differential operator, {\it and thus not  a tensor.} {But, as we will see, this is because we are considering this object defined in the full $3D$ Minkowski space. Let us pass to the dual object
\be
\mathcal{F}^{d} := \frac{1}{2}\epsilon^{dab}\mathcal{F}_{ab}\equiv -K^d + \epsilon^{dab} K_a \mathcal{A}_b + \epsilon^{dab} \mathcal{A}_a \mathcal{A}_b\;.
\ee
Then restricting this dual object (which is still not a tensor) to} the normal to the surface (\ref{EXaXaismnsn}) gives
\beqa\label{ProjField}
\mathcal{F} := X^a \mathcal{F}_{a} &= & \epsilon^{dab}X_d K_a \mathcal{A}_b + \epsilon^{dab}X_d \mathcal{A}_a \mathcal{A}_b\cr&&\cr&= & W^b \mathcal{A}_b + \epsilon^{dab}X_d \mathcal{A}_a \mathcal{A}_b\;,
\eeqa
where $W^a$ are the conformal Killing vectors (\ref{conKillers}), and we used (\ref{XFXidnt}).
{Using (\ref{Gauge}) it is easy to find how $\mathcal{F}_{ab}$ transforms
\be\label{FieldTransf}
\mathcal{F}_{ab} \mapsto \epsilon_{abc}K^c + U^{-1}\mathcal{F}_{ab} U\;.
\ee
Again, while (\ref{FieldTransf}) does not give the usual transformation of a field strength, it immediately leads to the correct transformation of the ``projected'' field strength (\ref{ProjField})
\be\label{PFieldTransf}
\mathcal{F} \mapsto U^{-1}\mathcal{F} U\;.
\ee
}\al{ We thereby have constructed  a field strength from the \al{embedding space} potentials ${\cal A}$.  We note that while ${\cal A}$ are vectors in three-dimensional Minkowski space, the field strength is a space-time scalar, rather than  being the typical  rank 2 antisymmetric  tensor.}

Next we construct dynamics using the degrees of freedom $ \mathcal{A}$. A natural choice  is to demand that the equations of motion lead to  zero field strength $\mathcal{F}=0$. In order to write down a gauge invariant action ${\cal S}$, we introduce a trace (tr) over $su(1,1)$ Lie-valued functions and a measure   $d\mu$  on the hyperboloid, which is consistent with the cyclic trace identity for the Poisson bracket, $\int d\mu\,\{{\cal A},{\cal B}\}{\cal C}=\int d\mu\,{\cal A}\{{\cal B},{\cal C}\}$. We choose the following action
\be
{\cal S}=-\frac 12\int d\mu\;{\rm tr} \mathcal{ A}_a X^a\left({\mathcal{\mathcal{F}}}+  \mathcal{ A}_b X^b\right)\label{JTactnSa}\;.
\ee
Variation with respect to ${\cal A}_a$ leads to the equations of motion
\be
X^a \mathcal{F}+\epsilon^{abc}K_b\left({\cal A}_dX^d X_c\right)-\epsilon^{abc}[{\cal A}_dX^d,{\cal A}_b] X_c+2{\cal A}_dX^d X^a =0 \;.\label{eomSfvrs}
\ee
Upon contracting with $X_a$ and using $\epsilon^{abc} X_a \{ X_b,X_c\}=2$, we get the desired result
\be
\mathcal{F}=0\;.\label{shFeqz}
\ee
The remaining tangential components state that ${\cal A}_b X^b$ is covariantly constant
\be
{\cal D}_a \left({\cal A}_b X^b\right):=K_a\left({\cal A}_b X^b\right)-[{\cal A}_b X^b,{\cal A}_a]=0 \;.\label{Tangent}
\ee

Eq. (\ref{shFeqz})  is solved by  pure gauges ${\cal A}=\mathcal{A}^{(0)},$ where
\be
\mathcal{A}^{(0)}_a[V]=V^{-1}(K_aV)=V^{-1}\{X_a,V\} \;,\qquad V\in SU(1,1)\;.\label{purgag}
\ee
We note that they are tangential to the hyperboloid,
\be
X^a \mathcal{A}^{(0)}_a[V]=0\;,\label{prgnoncom}
\ee
so $\mathcal{A}^{(0)}_a$ trivially satisfies (\ref{Tangent}), too.

We note that  (\ref{shFeqz}) can also have solutions whereby the normal components of ${\cal A}$ are nonvanishing, $X^a \mathcal{A}_a\ne 0$, which necessarily cannot be pure gauges.  {\it Thus, here zero field strength does not imply, even locally, that \al{ the embedding space potential   } is a pure gauge.}
For example,   consider
\be {\cal A}_a =i(X^b\tau_bX_a-\tau_a)\;,\label{ansol}\ee
 where
$\{\tau_a,\,a=0,1,2\}$, is a basis for the $su(1,1)$ Lie algebra, with
\be[\tau_a,\tau_b]=-2i\epsilon_{abc}\tau^c\;.\label{crsftau}\ee
It is easy to see that (\ref{ansol}) satisfies the zero curvature condition (\ref{shFeqz}).  Here one also gets that  $X^a \mathcal{A}_a=-2i X^a \tau_a$, and hence ${\cal A}$  cannot be written in the form (\ref{purgag}).  The covariant derivative of  $X^a {\cal A}_a$ does not vanish for this example, and hence the tangential components of the equation of motion (\ref{Tangent}) are not satisfied.

For pure gauge  solutions the on-shell action is unaffected by the addition of terms involving  arbitrary powers of  $X^a \mathcal{A}_a$ to the action (since $X^a \mathcal{A}_a$ vanishes for pure gauges).  So for example,   the action (\ref{JTactnSa}) can be generalized to
\be
{\cal S}'=-\frac 12\int d\mu\;{\rm tr} \mathcal{ A}_a X^a\left({\mathcal{\mathcal{F}}}+ c\, \mathcal{ A}_b X^b\right)\label{JTactnprime}\;,
\ee
where $c$ is an arbitrary real constant. The  equations of motion (\ref{eomSfvrs}) are generalized to
\be
X^a \mathcal{F}+\epsilon^{abc}K_b\left({\cal A}_dX^d X_c\right)-\epsilon^{abc}[{\cal A}_dX^d,{\cal A}_b] X_c+2c{\cal A}_dX^d X^a =0 \;,\label{eomSfvrs2}\ee
with the normal component now giving
\be {\cal F}=2(1-c){\cal A}_a X^a\;.\label{eomwcnz}\ee

\subsubsection{Formulation II}\label{Form2}
{Noting that $\epsilon^{dab}X_d K_a \mathcal{A}_b = \epsilon^{dab}K_a (X_d \mathcal{A}_b) - 2X^a \mathcal{A}_a$, the form of the field strength (\ref{ProjField}) suggests the following redefinition of the potentials, ${\cal A}_a\rightarrow A_a$,
\be
A_a := \epsilon_{adb} X^d \mathcal{A}^b - X_a (X^b \mathcal{A}_b)
\ee
and its inverse
\be
\mathcal{A}_a = -\epsilon_{abc} X^b {A}^c - X_a (X^b {A}_b)\;,
\ee
with $X^b \mathcal{A}_b = X^b {A}_b$ due to (\ref{EXaXaismnsn}). Then with the help of (\ref{XFXidnt}), we can write the projected field strength (\ref{ProjField}) in terms of the new gauge field
\be
\mathcal{F} = - K^a A_a + \epsilon^{abc}{ A}_a { A}_b X_c - 2 X^a A_a\;.
\ee
Once again,  $X^b \mathcal{A}_b = X^b {A}_b$ transforms in the adjoint representation of the gauge group, while the potentials $A_a$ transforms as
\be
{A}'_a = U^{-1} W_a U + U^{-1}{A}_a U\;,\label{straightgt}
\ee
where we used the definition (\ref{conKillers}). Then we can redefine our field strength as follows
\be
F := -(\mathcal{F} + 2 X^a \mathcal{A}_a ) = K^a A_a - \epsilon^{abc}{ A}_a { A}_b X_c\;.\label{straightF}
\ee
This still transforms in the adjoint representation of $SU(1,1)$.
}
 In comparing with the  formulation in \ref{FormI} where the gauge transformations were generated using the Killing vectors and the field strength was defined using the conformal Killing vectors, here we see from (\ref{straightgt}) and (\ref{straightF})  that the roles of the Killing vectors and conformal Killing vectors are reversed.
\al{ Once again, the embedding space potentials  yield a field strength scalar, rather than a rank 2 antiymmetric tensor. In the following subsection we will show that $F$ is the standard field strength for connections that are restricted to the two dimensional submanifold.}

A natural choice for the action in terms of the potentials $A_a$ is
\be
S_{JT}=-\frac 12\int d\mu\;{\rm tr}( { A}_a X^a\,{F} )\label{JTactn}\;.
\ee
Arbitrary variations in  $A_a$ lead to the  equations of motion\footnote{In both formulations we have assumed that  variations vanish at spatial infinity.  For the case of non vanishing variations $\delta A$ at the spatial boundary one picks up the boundary term $   -\int dX^a\, {\rm tr}{ A}_b X^b\delta A_a$. This variation may be eliminated by adding a term $   \int dX^a\, {\rm tr}{ A}_b X^b  A^{bnd}_a$ to the action, where $A^{bnd}_a $ represent boundary fields and ${ A}_b X^b$ is kept fixed at the boundary.}
\be
FX_a-\{X_a,{ A}_b X^b\}- \epsilon_{abc}[A^b,{ A}_d X^d]X^c=0
\label{eqfrAndB} \;.
\ee
Upon projecting out the normal component  from (\ref{eqfrAndB}) by contracting with $X_a$, we get that the field strength vanishes,
\be
F=0\label{stFeqz}\;.
\ee
In comparing with the previous formulation \ref{FormI}, and using (\ref{eomwcnz}) and (\ref{straightF}), this result is recovered from the action (\ref{JTactnprime}) when $c=2$.
  The equations of motion along the tangential directions once again state that the covariant derivative of $X^aA_a$ vanishes,
\be
D_a(A)\Big(X^bA_b\Bigr)=0\,,\,\label{DaXbAb}
\ee i.e., $X^aA_a$ is covariantly constant.  Here the covariant derivative is defined by
\be D_a(A)\,\Phi=\{X_a,\Phi\}+ \epsilon_{abc}[A^b,\Phi]X^c \;.\label{Cvrntdrv}
\ee

Pure gauge potentials are once again solutions to the zero curvature condition (\ref{stFeqz}).  Here they are expressed in terms of the conformal Killing vector
\be
A^{(0)}_a[V]=V^{-1}(W_aV)=\epsilon_{abc} X^b\,V^{-1}\{X^c,V\} \;,\qquad V\in SU(1,1)\label{purgag2}
\ee
and have zero field strength, $F=0$. Just as in the previous formulation  \ref{FormI}, they  are tangential to the hyperboloid,
\be
X^a A^{(0)}_a[V]=0\;.\label{prgnoncom}
\ee
The on-shell action for pure gauge solutions is once again unaffected by the addition of terms involving  arbitrary powers of  $X^a {A}_a$ to the action.  On the other hand, as we show in subsection \ref{recovery},  the  choice for the action in (\ref{JTactn}) defines  the standard  dynamics of JT gravity.

Just as in the previous formulation \ref{FormI}, the zero curvature condition does not imply (even locally) that the \al{embedding space} potential  must be a pure gauge.  A counter example is \be A^{\mbox{\tiny tran}}_a=-B X_a\;,\label{AislXa}\ee where $B$ is some non-zero Lie algebra-valued function on the hyperboloid.  From (\ref{straightF}) it easily follows that it is a solution to $F=0$.  Note that $X^a A_a$ doesn't vanish in this case.  If we try to write this solution as a pure gauge we get
\be
-B X_a=\epsilon_{abc} X^b\,V^{-1}\{X^c,V\}\;.
\ee
But then contracting  both sides with $X^a$  gives $B=0$.
The additional equation of motion, (\ref{Cvrntdrv}), is satisfied provided that $B $ is restricted to being  a constant on the hyperboloid. In that case (\ref{AislXa}) is a solution to the equations of motion   (\ref{eqfrAndB}).

More generally, the sum of a pure gauge (\ref{purgag2}) and transverse potential
the (\ref{AislXa}) also satisfies $F=0$. (Note in this regard that the transverse potential $A^{\mbox{\tiny tran}}$ does not contribute to the nonlinear term in $F$.) $ A=A^{(0)}+ A^{\mbox{\tiny tran}}$ is a solution to the equations of motion (\ref{eqfrAndB}) provided that
\beqa 0= D_a(A)\,B&=&\{X_a,B\}+ \epsilon_{abc}[A^{(0)b},B]X^c\cr&&\cr&=&\{X_a,B\}+ [V^{-1}\{X_a,V\},B]\cr&&\cr&=&V^{-1}\{X_a,VBV^{-1}\}V\;,\eeqa
which implies that $VBV^{-1}$ is a constant Lie algebra element on the hyperboloid. Though we have not proven this rigorously, we believe that $ A=A^{(0)}+ A^{\mbox{\tiny tran}}$  is the most general solution to the equations of motion (\ref{eqfrAndB}), which is supported by the consideration of the projected theory. \al{Below we will see that after  projecting formulation II to the two-dimensional submanifold, we recover the standard gauge theory formulation of JT gravity, whereby the zero field strength condition yields a pure gauge  directed along the hyperboloid (which is the projection of $A^{(0)}$), while $A^{\mbox{\tiny tran}}$  corresponds to the solution for the scalar field.}

\subsection{Recovering the standard formulation}\label{recovery}

In the standard formulation of JT gravity one assumes the existence of a
coordinate chart  on a  two dimensional domain, with some coordinates $x=(x^0,x^1)$. The field content  is commonly taken to be $su(1,1)$-valued potentials $a_\mu$, along with an $su(1,1)$-valued space-time scalar $b$ on the two dimensional domain.  $a$ and $b$ gauge transform according to
\be  a_\mu\rightarrow a'_\mu =U^{-1} \partial_\mu U + U^{-1}a_\mu U\; , \qquad b\rightarrow b' = U^{-1}b U\;,\label{littleab}
\ee
$\partial_\mu$  again denoting partial  derivatives  with respect to coordinates on the surface.  The field strength tensor is defined as usual by
\be f_{\mu\nu}=\partial_\mu a_\nu-\partial_\nu a_\mu+[a_\mu,a_\nu]\;,\ee
from which one constructs the invariant action
\be
S_{JT}=\frac 12\int d^2x\;\epsilon^{\mu\nu}\,{\rm tr}\,bf_{\mu\nu}\;.\label{actnltlab}
\ee
The resulting equations of motion are
\be
f_{\mu\nu}=0\;,\quad \partial_\mu b+[a_\mu,b]=0\;.\label{wontoo3}
\ee
General local solutions for the potentials are now \textit{only} pure gauges, which here take the  usual form \be a_\mu^{(0)}[V]=V^{-1}\partial_\mu V\;,\;\;\quad V\in SU(1,1)\label{pgponH2}\;,\ee
while solutions for  $b $ are $b=V^{-1}\ell V$, $\ell$ being an arbitrary constant element in the $su(1,1)$ Lie algebra.

We next show that the above  system is trivially recovered by  projecting the $SU(1,1)$ gauge theory of subsection \ref{Form2}   (formulation   II) to the two-dimensional  hyperboloid. We can then express the embedding coordinates in parametric form, $X^a=X^a(x)$, consistent with the constraint  (\ref{EXaXaismnsn}).
Any Poisson brackets on  the hyperboloid can be written locally as
\be \{{\cal G},{\cal H}\}=\frac  1{\sigma(x)}\epsilon^{\mu\nu}\partial_\mu{\cal G}\partial_\nu{\cal H}\;,\label{PBofGH}\ee for functions ${\cal G}(x) $ and  ${\cal H}(x) $. As in (\ref{indmtric}),
the function $\sigma(x)$ should be chosen to be compatible with the natural Poisson structure, (\ref{su11pba}), which means
\be\epsilon^{\mu\nu}\partial_\mu X^a\partial_\nu X^b={\sigma(x)}\,\epsilon^{abc} X_c\;,\label{1one5}\ee
or equivalently,
\be
\epsilon_{abc}\partial_\mu X^a\partial_\nu X^b=-{\sigma(x)}\, \epsilon_{\mu\nu}X_c\;.\label{dmuXadnuXb}
\ee
The function $\sigma(x)$ also enters in the expression for the integration measure \be d\mu=2\sigma(x) d^2x\;,\label{intmeas}\ee  as well as in  the Killing vectors and conformal Killing vectors. From
 (\ref{onedfKv}) and (\ref{conKillers}) we get
\be
K^a= \frac  1{\sigma(x)}\epsilon^{\mu\nu}\partial_\mu{X^a}\partial_\nu\;,\quad W_a= \frac  1{\sigma(x)} \epsilon^{\mu\nu}\epsilon_{abc}X^b\partial_\mu{X^c}\partial_\nu \;,\label{locisomgens}\ee
respectively.  Derivatives  on the surface are obtained by projecting the conformal Killing vectors along the tangent directions
\be
\partial_\mu X^a W_a=\partial_\mu\;.\label{W_deriv}
\ee
This is in contrast with the projection  of  the  Killing vectors along the surface
\be
\partial_\mu X^a K_a= \frac  1{\sigma(x)}\epsilon^{\rho\nu}g^{\mbox{\tiny ind}}_{\rho \mu}\partial_\nu\;.
\ee
Comparing (\ref{W_deriv}) and (\ref{straightgt}) immediately suggests that we should project the formulation II of the gauge theory in terms of potentials $A$, rather than  the formulation I of the gauge theory in terms of potentials ${\cal A}$.

Both $a_\mu$ and $b$ are recovered from the potentials in the three-dimensional background $A_a$.  By making the identification
\be
a_\mu(x)=A_a[X(x)]\,\partial_\mu X^a\;,\quad b(x)=A_a[X(x)]\, X^a(x)\;,\label{tan+norm}
\ee
we see that the gauge transformations (\ref{littleab}) follow from (\ref{straightgt}).  Moreover, the field strength (\ref{straightF}) is related to $f_{\mu\nu}$ by
\be F=-\frac 1{2\,\sigma(x)}\epsilon^{\mu\nu}f_{\mu\nu}\label{1ninten}\;,\ee
and the action (\ref{JTactn}) is identical to (\ref{actnltlab}).  \al{(\ref{1ninten}) shows that  $F$ indeed represents the field strength associated with connections on the surface.  Note that using (\ref{tan+norm}), the projection of  a pure gauge solution (\ref{purgag2}) for the embedding potentials $A$, yields a pure gauge for the potentials  $a_\mu$, while  the scalar field $b$ vanishes.  So the addition  of non-pure gauge solutions  (\ref{AislXa}) in the embedding space is essential for having solutions with non-trivial $b-$field in the projected theory.}

The potentials $a_\mu$ contain the gravitational degrees of freedom. More specifically, the zweibein and spin connection fields, $\{e^1_{\;\;\mu},e^2_{\;\;\mu}\}$  and $\omega_\mu=[\omega^1_{\;\;\,2}]_\mu$, respectively.  For this we re-introduce the $su(1,1)$ Lie algebra basis
$\{\tau_a,\,a=0,1,2\}$ with commutation relations (\ref{crsftau}).  We can take a $2\times 2$  matrix representation for $\tau_a$, and hence relate them to the Pauli matrices $\sigma_a$.  We choose  : $\;(\tau_0,\tau_1, \tau_2)=(\sigma_3,i\sigma_1,i\sigma_2)$, which then  leads to the identity
\be
\tau_a\tau_b=-i\epsilon_{abc}\tau^c-\eta_{ab}\BI\label{tauataub}\;.
\ee
 Let us now fix the definition of $SU(1,1)$ group elements by demanding $U$ satisfies
\be
U^\dagger \tau_0 U=\tau_0\;,   \quad {\rm det}\,U=1\;.
\ee
The zweibeine correspond to  the $\tau_1$ and $\tau_2$ components of  $a_\mu$, while the spin connection is associated with the $\tau_0$ direction.  Writing the expression in terms of one forms we have
\be a= a_\mu dx^\mu=-\frac i2\Bigl(e^1\tau_1+e^2\tau_2-\omega\tau_0\Bigr)\;,\label{zwibi}\ee
where $e^i=e^i_{\;\;\mu}dx^\mu$, $i=1,2$, and $\omega=\omega_\mu dx^\mu$.
 The $su(1,1)-$valued  curvature two form
$ f=da+a^2$
 contains the spin curvature $R=d\omega $ and the  torsion, $ T^i=de^i+\omega^i_{\;\;\,j} e^j\,$. (The indices $i,j=1,2$ are raised and lowered using the two-dimensional Euclidean metric.) Upon substituting (\ref{zwibi}) one has
\beqa
f&=&-\frac i2\Bigl(de^1\tau_1+de^2\tau_2-d\omega\tau_0\Bigr)-\frac 12\Bigl(e^1e^2\tau_1\tau_2-e^1\omega\tau_1\tau_0-e^2\omega\tau_2\tau_0\Bigr)\;\cr&&\cr&=& -\frac i2\Bigl(T^1\tau_1+T^2\tau_2-(R +e^1 e^2)\tau_0\Bigr)\;,\label{twnte6}\eeqa
where we used (\ref{tauataub}).\footnote{The choice of $\tau$-matrices was made to be consistent with the property $\omega^2_{\;\;\,1}=-\omega^1_{\;\;\,2}$, which applies for the case of  an   Euclidean  signature.}  Then from the equation of motion $f=0$ (\ref{wontoo3}),
\be T^1=T^2=R +e^1 e^2=0\;,\label{notorsion}\ee which defines $AdS_2$.
The metric tensor constructed from zweibein one forms,  which are obtained from the pure gauge, is given by
\be
g_{\mu\nu}[V]= e^1_\mu e^1_\nu+e^2_\mu e^2_\nu\;.\label{geiei}
\ee
The metric tensor depends on the choice for the $SU(1,1)$ group element.
While $g_{\mu\nu}[V]$, if nonsingular, describes $AdS_2$ it obviously is not in general the induced metric (\ref{indmtric}).  The question of which $V$ in $SU(1,1)$ yields the induced metric will be addressed in the following section.

Pure gauges (\ref{pgponH2})  that give singular metrics cannot be regarded as physical.  For instance, global, i.e. $X^a-$independent, $ SU(1,1)$ group elements (including the identity) give a zero metric tensor. Call ${\cal G}^{\mbox{\tiny phys}}$  the set of  $V\in SU(1,1)$  whose corresponding zweibein are such that det$\,e^a_{\;\;\mu}$ is nonzero and finite. ${\cal G}^{\mbox{\tiny phys}}$ does not form a group, since for example it does not contain the identity.
We can transform from any element $V\in{\cal G}^{\mbox{\tiny phys}}$  to any $W\in{\cal G}^{\mbox{\tiny phys}}$ using the right action  of
$g=V^{-1}W$ on  $V$.  The set ${\cal G}$ of all such elements $g$ (which  includes the identity) can be regarded as the gauge group as, at least locally, it provides a map between different $AdS_2$ metrics.

Local rotations of the zweibein one forms, $e^i\rightarrow {e'}^i=\Lambda^{ij} e^j $, $\Lambda^{ij}\Lambda^{ik}=\delta^{jk}$, leave the metric tensor (\ref{geiei}) unchanged. They are implemented by the right action of $U(1)$ on $V\in{\cal G}^{\mbox{\tiny phys}}$, $V\rightarrow V'=V e^{i\lambda \tau_0}$, where $\lambda$ is a function on the two-dimensional surface.  Thus $g_{\mu\nu}[Ve^{i\lambda \tau_0}]=g_{\mu\nu}[V]$.   The set of  $e^{i\lambda \tau_0}$
is a $U(1) $ subgroup of ${\cal G}$, which  serves as a little group, as it preserves the metric tensor. We can then argue, at least locally, that  gauge transformations can be decomposed into a product of  $U(1) $ gauge transformations and diffeomorphisms.

As stated above, the metric tensor  (\ref{geiei}) obtained from a pure gauge is not, in general,  the induced metric (\ref{indmtric}), which will be illustrated using different examples  in the next section. Moreover, ${\cal L}_{K^a}\,g_{\mu\nu}[V]$,  with $K^a$ given by (\ref{onedfKv}), need not vanish.  On the other hand, since  $g_{\mu\nu}[V]$, for any $V\in{\cal G}^{\mbox{\tiny phys}}$,  describes $AdS_2$ it must, at least locally,  be related to   $ g^{\mbox{\tiny ind}}_{\mu\nu}$ by a diffeormorphism.  Therefore, $ g_{\mu\nu}[V]$ possesses all the isometries of $AdS_2$, but its Killing vectors, as well as conformal Killing vectors, will not, in general, be given by (\ref{onedfKv}) and (\ref{conKillers}), respectively.

\section{Solutions}
\setcounter{equation}{0}

Here we consider  pure gauge solutions constructed from some  $V\in{\cal G}^{\mbox{\tiny phys}}$, and their corresponding  metric tensors  (\ref{geiei}).  While the latter all describe  $AdS^2$, at least locally, and  are invariant under the $SO(2,1)$ isometry group, the  isometry transformations can differ.  So we can ask: For which pure gauge solutions are the isometry transformations for (\ref{geiei}) the same as those for  the induced metric?  More specifically, from which element $ V^{\mbox{\tiny ind}}\in{\cal G}^{\mbox{\tiny phys}}$ do we recover the induced metric   (\ref{indmtric})?  We know that the answer is not unique, since  $\{ V^{\mbox{\tiny ind}} e^{i\lambda \tau_0}\}$ forms an equivalence class for  the induced metric tensor. Since the induced metric tensor is invariant under the action of the isometry group ${\cal I}$ generated by the Killing vectors (\ref{onedfKv}), one would expect that the answer for $ V^{\mbox{\tiny ind}}$  displays a high degree of symmetry.
 The requirement that  $ V^{\mbox{\tiny ind}}$ be invariant under the action of ${\cal I}$   is obviously  too strong, as the only invariant that can be constructed from the embedding coordinates is (\ref{EXaXaismnsn}).  Demanding that    $ V^{\mbox{\tiny ind}}$  is invariant under the action of the global $SU(1,1)$ group $G$ generated by $\tau_a$ is  also too strong.  (Here  $G$ is implemented using the adjoint action, $V\rightarrow g_0^{-1} V g_0$, $ g_0\in G$.) On the other hand, we can construct nontrivial  $SU(1,1)$ group elements which are invariant under the action of the diagonal subgroup $({\cal I} \otimes G)_{diag}$.

 Below we construct  two classes of pure gauge potentials. (We set $A_aX^a={\cal A}_aX^a=0$.) In the first, considered in  subsection \ref{Maxsym}, $V$ is invariant under all of  $({\cal I} \otimes G)_{diag}$, and in the second, considered in subsection \ref{U1subgrp}, it is  invariant only under the action of the $U(1)$ subgroup of  $({\cal I} \otimes G)_{diag}$.  We shall show that $ V^{\mbox{\tiny ind}}$ is contained in the second class of solutions, and so it does not posses the full symmetry  $({\cal I} \otimes G)_{diag}$.
No coordinate charts are  required in writing down the results for $V$.

\subsection{Maximally invariant solutions}\label{Maxsym}

Invariance of $V$ under $({\cal I} \otimes G)_{diag}$ means that
\be
\{X_a,V\} +\frac i2 [\tau_a,V]=0\;,\quad a=0,1,2\;.
\ee
This  property is satisfied for any function of
\be
\gamma=X^a\tau_a\;,\label{gamma}
\ee
which satisfies
\be
\gamma^2=\BI\label{gamma2}\;.
\ee
We can then write down the following  $({\cal I} \otimes G)_{diag}$ invariant
\be V=e^{-i\theta\gamma}\;,\ee
where $\theta\;,\;0\le\theta<2\pi$, is an arbitrary parameter. From (\ref{gamma2}), we then have
\be
V=\cos\theta\,\BI+i\gamma\sin\theta\;.
\ee
It satisfies det $V=1$, and ${V}^\dagger\tau_0 V=\tau_0 $, and so is an $SU(1,1)$ group element.

From (\ref{purgag}), the corresponding pure gauge potentials ${\cal A}$ of formulation I are
\be {\cal A}_a=-i\sin\theta\cos\theta\, \epsilon_{abc}X^b\tau^c+i\sin^2\theta\,(\tau_a+X_a\gamma)\;,\ee
while in terms of the gauge potentials $A$ of formulation II one gets
\be
A_a=i\sin^2\theta\, \epsilon_{abc}X^b\tau^c+i\sin\theta\cos\theta\,(\tau_a+X_a\gamma)\;.\label{sol1A}
\ee
Using  (\ref{pgponH2})  the projection of the latter  gives the following $su(1,1)$ connection one forms on the surface
\be
a_\mu dx^\mu=-i\sin^2\theta\, \epsilon_{abc}X^a dX^b\tau^c+i\sin\theta\cos\theta\,\tau_a dX^a\;.
\ee
From (\ref{zwibi}) the corresponding zweibein and spin connection one forms are
\beqa
\frac 12 \,e^1&=&-\sin^2\theta\, (X^2dX^0-X^0dX^2) -\sin\theta\cos\theta\, dX^1\;,\cr&&\cr
 \frac 12 \,e^2&=&-\sin^2\theta\, (X^0dX^1-X^1dX^0) -\sin\theta\cos\theta\, dX^2\;,
\cr&&\cr
 \frac 12 \,\omega&=&-\sin^2\theta \,(X^1dX^2-X^2dX^1) +\sin\theta\cos\theta\, dX^2\;.
\eeqa
The resulting $AdS_2$ metric has a complicated form
$$g_{\mu\nu}[V]\;=\;4\sin^4\theta\;\;\times\qquad\qquad\qquad\qquad\qquad\qquad\qquad\qquad$$
$$\biggr\{\left(X^2\partial_\mu X^0-X^0\partial_\mu X^2+\cot\theta\, \partial_\mu X^1\right)\left(X^2\partial_\nu X^0-X^0\partial_\nu X^2+\cot\theta\, \partial_\nu X^1\right)$$
\be +\left(X^0\partial_\mu X^1-X^1\partial_\mu X^0+\cot\theta\, \partial_\mu X^2\right)\left(X^0\partial_\nu X^1-X^1\partial_\nu X^0+\cot\theta\, \partial_\nu X^2\right)\biggr\}\;.
\label{tf8}\ee
Of course, the answer is singular for $\sin\theta=0$.
For no value of $\theta$ is  the   metric tensor (\ref{tf8}) equal to the induced metric tensor  (\ref{indmtric}).  Moreover, even though $V$ posses the full $({\cal I} \otimes G)_{diag}$ symmetry,  ${\cal L}_{K^a} g_{\mu\nu}\ne 0$ for all $a=0,1,2$.  On the other hand, since $g_{\mu\nu}$ describes $AdS_2$, there exist another set of  Killing vectors ${K}_{(\theta)}^a,$ $a=0,1,2$, whose  Lie derivatives acting on $g_{\mu\nu}$ vanish.

\subsection{$U(1)$ invariant solutions}\label{U1subgrp}

Next we construct a family  of solutions which are  invariant only under the action of the $U(1)$ subgroup of $({\cal I} \otimes G)_{diag}$. The subgroup simultaneously rotates $X^1$ and $X^2$, and $\tau_1$ and $\tau_2$.  A  $U(1)$ invariant is
\be
\varrho=X^1\tau_1+ X^2\tau_2\;.
\ee
It satisfies $\{X_0,\varrho\} +\frac i2 [\tau_0,\varrho]=0\,$.
Upon introducing functions $f(X^0)$ and $h(X^0)$,  we can write  down the   following general expression for a $U(1)$ invariant
 $SU(1,1)$ group element
\be
V=\exp\left\{\frac{-i f(X^0)\,\varrho}{\sqrt{(X^0)^2-1}}\right\}e^{ih(X^0)\tau_0}\;.
\ee
Since the factor $e^{ih(X^0)\tau_0}$ does not affect the metric tensor  we set $h(X^0)=0$.
Using $\varrho^2=-\left((X^0)^2-1\right)\BI$, we get
\be
V=\cosh \left( f(X^0)\right)\BI-\frac{i\varrho}{\sqrt{(X^0)^2-1}}\sinh \left( f(X^0)\right)\;.\label{VUone}
\ee
$V$  satisfies det $V=1$, and ${V}^\dagger\tau_0 V=\tau_0 $, and so is an $SU(1,1)$ group element.

From (\ref{purgag}) the corresponding pure gauge potentials  $A$ of subsection \ref{Form2} are
\beqa
A_0&=&i\sqrt{(X^0)^2-1}\, f'\,\varrho\;,\cr&&\cr A_i&=&-i\tau_i \frac{\sinh f\cosh f}{\sqrt{(X^0)^2-1}}+i\epsilon_{0ij} X^j\tau_0\frac{\sinh^2f}{(X^0)^2-1}\cr&&\cr&&\qquad-\;i \frac{X^i\varrho}{\sqrt{(X^0)^2-1}}\left(X^0 f'-\frac{\sinh f\cosh f}{{(X^0)^2-1}} \right)\;,\quad i,j,...=1,2\;,\cr&&\label{A0AiU1}\eeqa
where the prime denotes a derivative with respect to $X^0$.  The corresponding $su(1,1)$ connection one form on the surface (\ref{pgponH2}) is
\beqa a_\mu dx^\mu&=&i\left(\frac{ {X^0}\sinh f\cosh f}{{(X^0)^2-1}}- f' \right)\frac{\varrho\, dX^0}{{\sqrt{(X^0)^2-1}}}
\cr&&\cr&&
-i\,\frac{\sinh f\cosh f}{\sqrt{(X^0)^2-1}}\tau_i dX^i-i\,\frac{\sinh^2f}{(X^0)^2-1}\epsilon_{0ij} X^idX^j\tau_0\,,\eeqa
from which we get the following zweibein and spin connection one forms:
\beqa
e^i&=&\frac {\sinh(2 f)}{\sqrt{(X^0)^2-1}}\left( dX^i - {\cal H}(X^0){X^i dX^0}\right)\;,\cr&&\cr
\omega&=&-\frac{2\sinh^2f}{(X^0)^2-1}\epsilon_{0ij} X^idX^j\;,
\eeqa
where
\be
{\cal H}(X^0)=\frac{ {X^0}}{{(X^0)^2-1}}-\frac {2 f' }{\sinh(2 f)}\;.\label{calF}
\ee
The resulting metric   for arbitrary $ f(X^0)$ is nontrivial:
\beqa g_{\mu\nu}[V]&=&\frac {\sinh^2(2 f)}{{(X^0)^2-1}}\,
\left( \partial_\mu X^i - {\cal H}(X^0){X^i \partial_\mu X^0}\right)\,\left( \partial_\nu X^i - {\cal H}(X^0){X^i \partial_\nu X^0}\right)\cr&&\cr
&=&\frac {\sinh^2(2 f)}{{(X^0)^2-1}}\,
\Bigl( \partial_\mu X^i \partial_\nu X^i + {\cal H}(X^0)^2X^i X^i \partial_\mu X^0\partial_\nu X^0 \cr&&\cr&&\qquad\quad\qquad
-\; {\cal H}(X^0)X^i \left(\partial_\mu X^0 \partial_\nu X^i +\partial_\nu X^0 \partial_\mu X^i \right)\Bigr)\cr&&\cr
&=&\frac {\sinh^2(2 f)}{{(X^0)^2-1}}\,
\bigg( - \left(\frac {(X^0)^2}{\left((X^0)^2-1\right)^2}-\frac {(2f' )^2 }{\sinh^2(2 f)}\right)((X^0)^2-1)\,\partial_\mu X^0 \partial_\nu X^0 \cr&&\cr&&\qquad\qquad\qquad\qquad\qquad+\;\;\partial_\mu X^i \partial_\nu X^i \,\biggr)\;,
\eeqa
where we used (\ref{EXaXaismnsn}) and (\ref{calF}).
However, for the special choice
\be  \cosh(2f)=X^0\;,\label{ftgim}\ee
we  get   $\frac {2 f' }{\sinh(2 f)}=\frac 1{(X^0)^2-1}$,
and  then  recover the induced metric, $ g_{\mu\nu}[V]=g^{\mbox{\tiny ind}}_{\mu\nu}\,$.
The $SU(1,1)$ group element (\ref{VUone}) simplifies in this case to
\be
V^{\mbox{\tiny ind}}=\frac 1{\sqrt 2}\left(\sqrt{X^0+1}\;\BI-\frac{i\varrho}{\sqrt{X^0+1}}\right)\;,
\ee  while the $su(1,1)$ gauge potentials (\ref{A0AiU1}) reduce  to
\beqa
A^{\mbox{\tiny ind}}_0&=&\frac i 2\,\varrho\;,\cr&&\cr A^{\mbox{\tiny ind}}_i&=&-\frac i2\left(\tau_i + \frac{X_i\varrho-\epsilon_{0ij} X^j\tau_0}{X^0+1}\right)\;.
\eeqa
In terms of the potentials  ${\cal A}$ of formulation I (discussed in subsection \ref{FormI} ) one gets
\beqa
{\cal  A}^{\mbox{\tiny ind}}_0&=&\frac i2 \left(\epsilon_{0ij} X^i\tau^j+\left(X^0-1\right)\tau_0\right)\;,\cr&&\cr {\cal A}^{\mbox{\tiny ind}}_i&=&-\frac i2\epsilon_{ij0}\left(X^0\tau^j+\frac{2X^0+1}{X^0+1}X^j\varrho\right)-\frac i2\frac {X^0X_i}{X^0+1}\tau_0\;.
\eeqa

Once again, $V^{\mbox{\tiny ind}}$ is not unique, as there  is equivalence class ${\cal V}^{\mbox{\tiny ind}}=\{ V^{\mbox{\tiny ind}} e^{i\lambda \tau_0}\}$  of $SU(1,1)$ group elements
that give the induced metric tensor.   One element in particular,  ${ V}^{\mbox{\tiny ind}}\, e^{-\frac{i\pi}4 \tau_0}$, has the property that it provides a map from $\tau_0$ to the  $({\cal I} \otimes G)_{diag}$ invariant $\gamma$, (\ref{gamma}).\footnote{In \cite{Pinzul:2021cjz} this  map was  used to transform from the covariant Dirac operator on $AdS_2$ to a noncovariant form of the Dirac operator.} This follows from the identity
\be\label{sigmatogamma}
\left( V^{\mbox{\tiny ind}}\,e^{-\frac{i\pi}4 \tau_0}\right)^{-1}\tau_0\left( V^{\mbox{\tiny ind}}\, e^{-\frac{i\pi}4 \tau_0}\right)=\gamma\;.
\ee
From the space-time point of view, we can think of (\ref{sigmatogamma}) as the transitive action of $SU(1,1)$ on the fixed point $X=(1,0,0)$, which is stable under the $U(1)$ action. The image of the transitive action will be the whole $AdS_2$. This motivates the most ``natural'' construction  of pure gauge potentials leading to the induced metric: Find an $SU(1,1)$ element that takes $\tau_0$ to $\gamma$, as in (\ref{sigmatogamma}), and obtain from it the pure gauge solution. Of course, one still should verify that it will lead to the induced metric, but this will be a much simpler check.

\section{Non-commutative JT gravity}\label{SectionNCfree}

While we saw that formulation II of the  gauge theory  (described in subsection \ref{Form2}) was useful for recovering the standard formulation of JT gravity, formulation I of subsection \ref{FormI} offers a more natural starting point for writing down the generalization of the  gauge theory to a non-commutative space.  Here we briefly outline the procedure.  We shall be concerned with the  well known isometry preserving non-commutative deformation $NCAdS_2$ of  the $AdS_2$ background.\cite{Ho:2000fy,Fakhri:2011zz,Jurman:2013ota,Stern:2014aqa,Chaney:2015ktw}  Scalar and fermionic field theories have been examined on  $NCAdS_2$ in \cite{Pinzul:2017wch,deAlmeida:2019awj,Lizzi:2020cwx,Pinzul:2021cjz,Pinzul:2022pmg,Stern:2022zwo}.   Below after first reviewing the noncommutative deformation of $AdS_2$, we give the construction of the  gauge theory on  $NCAdS_2$. There is a well known obstacle to preserving the determinant of the gauge group on a noncommutative space.  With that in mind we extend the gauge group from $SU(1,1)$ to $U(1,1)$. Noncommutative gravity was previously explored in \cite{Cacciatori:2002ib}, where a description was given in terms of coordinate charts and star products.   In line with the central theme of this article,
our procedure avoids the  use of coordinate charts or star products.

The deformation of $AdS_2$ to $NCAdS_2$ introduces a real parameter $\alpha$, the noncommutativity parameter, $\alpha\rightarrow 0 $ defining the commutative limit. The deformation to  $NCAdS_2$ replaces the embedding coordinates  $X^a$ by hermitian operators,  $\hat X^a$, with the condition (\ref{EXaXaismnsn}) becoming the constraint
\be
\hat X^a \hat X_a=-\BI\;,\ee while the Poisson brackets (\ref{su11pba}) are replaced by the commutators
\be\label{adstoocrs}
[\hat X^a,\hat X^b]=i\alpha\,\epsilon^{abc}\hat X_c\; ,
\ee
where  $\BI$ is the identity operator.   A non-commutative version of the Killing vectors $\hat K_a$ can be introduced which preserves the $su(1,1)$ algebra (\ref{so21Kil}).\cite{Pinzul:2017wch}  $\hat K_a$  is an inner derivative,  its action  on functions ${\cal G}(\hat X)$ of $\hat X^a$  defined by
\be
\hat K^a{\cal G}(\hat X)=\frac 1{i\alpha}[\hat X^a,{\cal G}(\hat X)]\;.
\ee
This leads to the non-commutative analogue of (\ref{KaonXb}).\footnote{{We note, in contrast, that the noncommutative extension  of the conformal Killing vectors is not  obvious.  Any such extension may not preserve   the commutation relations  (\ref{othrctrs}), nor, for that matter, be consistent with the Leibniz rule.  This is consistent with the breaking of conformal invariance  by noncommutativity. In \cite{Pinzul:2017wch,deAlmeida:2019awj,Lizzi:2020cwx} it was shown that conformal symmetry is recovered only as one approaches the asymptotic boundary.}}

A standard  way to write down  a noncommutative gauge theory is based on the introduction of so-called covariant coordinates  $\hat Y^a$.  Here they are defined to be $u(1,1)$ Lie algebra-valued, and  transform under the adjoint action of  the gauge group, which once again, should be extended to $U(1,1)$.
Call  $\hat U$ a   $U(1,1)-$valued  function  of $\hat X^a$. Thus $\hat Y^a$   gauge transform according to
\be
\hat Y^a \rightarrow \hat Y^{'a}=\hat U^{-1}\hat Y^a\hat U\;.\label{ncvrtnYa}
\ee  Next express
$\hat Y^a$  in terms
 of the noncommutative analogues $\hat{ \cal A}$ of the  potentials ${\cal A}^a$
of subsection \ref{FormI}, which we now consider as being valued in the $u(1,1)$ Lie algebra.  We write
\be\hat Y^a=\hat X^a+i\alpha\hat{ \cal A}^a\;,\label{rdfofY}\ee
Then from (\ref{ncvrtnYa}),
$\hat{ \cal A}^a$ transform according to
\be
\hat{ \cal A}^a\rightarrow {\hat{ \cal A'}}^a=\hat U^{-1}\hat K^a\hat U+\hat U^{-1}\hat{\cal A}^a\hat U\;,
\ee
which is the noncommutative version of the gauge transformation (\ref{Gauge}).

From $Y^a$ we can construct a couple of space-time scalars  which transform covariantly under the action of the gauge group.  One is
\be [\hat Y]^2:=\hat Y^a\hat Y_a=-\BI +i\alpha \left(\hat X^a\hat{\cal  A}_a+\hat{\cal  A}_a \hat X^a\right)-\alpha^2\hat{\cal  A}_a\hat{\cal  A}^a\;,\label{NCB}\ee
which contains $X^a{\cal  A}_a$ in the commutative limit $\alpha\rightarrow 0$.  More precisely, $X^a{\cal  A}_a$ is the commutative limit of $-\frac i{2\alpha}\left(\BI+[\hat Y]^2\right)$.
Another
 contains  the field strength (\ref{ProjField}) in the commutative limit.  It is
\be [\hat Y]^3:=\epsilon^{abc}{ \hat Y}_a {\hat  Y}_b \hat Y_c\;=\;i\alpha \BI +\frac {1}2\alpha^2\left(\hat{\cal A}_c\hat X^c+\hat X^c\hat{\cal A}_c
-\hat{\cal F}^c\hat Y_c-\hat Y_c\hat{\cal F}^c\right)
\label{NCcurv}\;,\ee
where
\be
\hat {\cal F}^c=\epsilon^{abc}\left(\hat K_a\hat{\cal A}_b+\hat{\cal A}_a\hat{\cal A}_b\right)\;.
\ee
In the commutative limit it reduces to
$\; i\alpha  +\alpha^2\left(X^a{\cal A}_a-{\cal  F}\right)\,,$ with  ${\cal  F}$ defined in (\ref{ProjField}).   [In terms of the field strength $F$ of Formulation II, (\ref{NCcurv}) reduces to $\; i\alpha  +\alpha^2\left(F+3X^a{ A}_a\right)\,.]$  Using both
(\ref{NCB}) and (\ref{NCcurv}) we see that ${\cal F}$ is the commutative limit of
\be
-\frac 1{\alpha^2}[\hat Y]^3 +\frac i{2\alpha}\left(\BI-[\hat Y]^2\right)\;.
\ee

A gauge  invariant action $\hat S$ can now be constructed from   (\ref{NCB}) and (\ref{NCcurv}) which  reduces to  (\ref{JTactnSa}) in the commutative limit.  Introducing  an invariant trace Tr over the algebra, one has
\beqa \hat S&=&\frac{i\alpha}4\,{\rm Tr}\left(\BI+[\hat Y]^2\right)\left(-\frac 1{\alpha^2}[\hat Y]^3 +\frac i{2\alpha}\left(\BI-[\hat Y]^2\right)-\frac i{2\alpha}\left(\BI+[\hat Y]^2\right)\right)\cr&&\cr&=&-\frac i{4\alpha}\,{\rm Tr}\left(\BI+[\hat Y]^2\right)\left([\hat Y]^3 + i{\alpha}[\hat Y]^2\right)\label{NCJTactn}\;.\eeqa
Of course, this answer is not unique as the inclusion of any higher order correction terms in $\alpha$ will  also   reduce to  (\ref{JTactnSa}) in the commutative limit. Equations of motion are obtained by varying $\hat{\cal A}_a$.  Their expressions are rather involved and the solutions are not transparent, which will not be pursued here.

\section{Concluding remarks}
\setcounter{equation}{0}

In this paper, we studied the embedding space approach to  pure, i.e., without matter, JT gravity.   \al{The standard gauge theory formulation was recovered from the mapping (\ref{tan+norm}), where the normal components of the embedding space potentials  are mapped to the scalar field. Furthermore, from the action (\ref{JTactn}) one recovers the standard action (\ref{actnltlab}), indicating an equivalence with the standard gauge theory formulation. Although the considerations in our paper are purely classical, due to established equivalence with the standard gauge theory formulation, we do not expect that the quantized version of our system will differ from that of the standard formulation.  For example, in the path integral approach the integration over the transverse component of $A_a$ must be identical to the integration over the scalar field, and it should  enforce  the constraint $F=0$ (for formulation II).  Nevertheless, a careful analysis of the quantum theory is warranted.} 

 Below we outline  other  possibilities for extensions and generalizations of this work that can be considered \al{within a classical framework}.

While the treatment of JT gravity presented here  assumed a Euclidean signature, there are no formal obstacles to replacing the latter with a Lorentzian signature. The first step in this regard is to change  the background metric  (\ref{Ebkgrnd}) to, say ${\rm diag}(-+-)$, while keeping  (\ref{EXaXaismnsn}). In either the Euclidean or Lorentzian version of the theory, the embedding coordinates provide for a global description of the theory. On the other hand, local coordinates  play an important role for the Lorentzian theory, as they define the causal structure, and so field theories written on different coordinate patches can have inequivalent vacua. A physically non-trivial example of a local coordinate patch on the hyperboloid is  the so-called $2d$ black hole.\cite{Spradlin:1999bn,AP2015} It is somewhat analogous to the Rindler wedge, in that it possesses horizons, but no space-time singularities. Another, more complicated, question related to going to the Lorentzian signature is that in order to avoid closed time-like curves, one has to decompactify the embedded surface. In this case, the role (and the very possibility) of our formulation in terms of the embedding space  requires further studies.

Matter fields and interaction terms  can be introduced to the system. Also, the theory seems to be very flexible for various natural generalizations that may not be so obvious in the standard two-dimensional formulation. For example, additional terms involving $b=A_aX^a={\cal A}_aX^a$ \al{which are allowed by the symmetry constraints will modify the equations of motion, which will possibly lead to new types of solutions.  We hope to address this issue in future works.}

{A detailed analysis of the boundary dynamics   is called for.  In fact, if  our gauge theory is truly equivalent to the metric formulation of JT gravity then it should be possible to recover a Schwarzian action on the boundary.\cite{MSY2016} Demonstrating this might represent a challenge in our coordinate free approach, as even for the standard gauge theory formulation (although the  Schwarzian action was successfully recovered) the derivation requires some nontrival steps.\cite{Mertens:2018fds,Blommaert:2018oro,Iliesiu:2019xuh,Ferrari:2020yon}.}

Here we have only outlined the procedure for going to a noncommutative version of JT gravity. Clearly,  much  work should be done in understanding both its mathematical and physical aspects. For example, the system may have a description in terms of unitary representations of $U(1,1)$.  Also, the role of the now non-decoupled extra $U(1)$ field should be clarified.

The treatment of JT gravity presented here could have utility in connection to duality, and more specifically to the SYK model.\cite{MSY2016}   In particular, the noncommutative version of this theory has the potential of incorporating quantum gravity effects, verifying the validity of the AdS/CFT correspondence beyond the classical (super)gravity regime. It would also be interesting to study the relation, if any, to our previous works in this direction (see \cite{Pinzul:2022pmg,Stern:2022zwo} and the references therein).

Finally, the coordinate-free procedure developed here for writing a gauge theory on  $AdS_2$ should be generalizable to other isometric  manifolds  immersed  in  fixed backgrounds.  In this paper, the number of isometries and background dimensions agreed. More generally, the choice of isometries and background space cannot be arbitrary.

\bigskip
\bigskip
{\large{\bf Acknowledgement}}

\noindent
A.P. acknowledges the partial support of CNPq under the grant no.312842/2021-0.

\bigskip
\bigskip

\appendice{$\quad${\large{\bf Killing vectors and conformal Killing vectors}}

\bigskip
Here we verify equations (\ref{Lieong}) and (\ref{rsclmetrc}).

{The Killing equation is
\be\label{Killing}
{\cal L }_{K^a}g_{\mu\nu}=K^{a\lambda}\partial_\lambda g_{\mu\nu}+(\partial_\mu K^{a\lambda} )g_{\lambda \nu}+(\partial_\nu K^{a\lambda} )g_{\mu\lambda }=0 \ ,
\ee
where, in our case, $K^{a\mu}$ is defined by (\ref{onedfKv}), i.e., $K^{a\mu}\partial_\mu \mathcal{G} = \{X^a , \mathcal{G}\}$ for any function $\mathcal{G}(X)$. Applying this to the induced metric (\ref{indmtric}) we get
\be\label{Killing1}
K^{a\lambda}\partial_\lambda g^{\mbox{\tiny ind}}_{\mu\nu} = \eta_{bc} \{X^a , \partial_\mu X^b \}\partial_\nu X^c + \eta_{bc} \{X^a , \partial_\nu X^c \}\partial_\mu X^b \;.
\ee
Combining the first term in (\ref{Killing1}) with the second term in (\ref{Killing}), we get
\beqa
&&\eta_{bc} \{X^a , \partial_\mu X^b \}\partial_\nu X^c + \partial_\mu K^{a\lambda} g^{\mbox{\tiny ind}}_{\lambda \nu} = \eta_{bc} \left( K^{a\lambda}\partial_\lambda \partial_\mu X^b \partial_\nu X^c + \partial_\mu K^{a\lambda} \partial_\lambda X^b \partial_\nu X^c\right)\nonumber\\
&&= \eta_{bc} \partial_\mu (K^{a\lambda} \partial_\lambda X^b) \partial_\nu X^c \equiv \eta_{bc} \partial_\mu \{ X^a , X^b \} \partial_\nu X^c = \epsilon^a_{\ cd} \partial_\mu X^d \partial_\nu X^c \;.
\eeqa
For the combination of the second term in (\ref{Killing1}) with the third term in (\ref{Killing}), we get the same expression with $\mu \leftrightarrow \nu$, so we arrive at the claimed result  (\ref{Lieong}).\footnote{Note that in our derivation we did not use the constraint (\ref{EXaXaismnsn}). So, the result holds in every case when the Killing action is given by (\ref{onedfKv}) with some Lie-type constant Poisson structure. This is not so for the case of the conformal Killings, see below.}

}
{For the case of the conformal Killing vector given in (\ref{conKillers}), we get
\beqa
{\cal L }_{W^a}g^{\mbox{\tiny ind}}_{\mu\nu}&=& W^{a\lambda}\partial_\lambda g^{\mbox{\tiny ind}}_{\mu\nu}+(\partial_\mu W^{a\lambda} )g^{\mbox{\tiny ind}}_{\lambda \nu}+(\partial_\nu W^{a\lambda} )g^{\mbox{\tiny ind}}_{\mu\lambda }\cr&&\cr&=&\epsilon^{abc}\left(X_b K^{\lambda}_c\partial_\lambda g^{\mbox{\tiny ind}}_{\mu\nu}+\partial_\mu (X_b K^{\lambda}_c )g^{\mbox{\tiny ind}}_{\lambda \nu}+\partial_\nu (X_b K^{\lambda}_c)g^{\mbox{\tiny ind}}_{\mu\lambda }\right)\cr&&\cr
&=&\epsilon^{abc}\left(X_b(  {\cal L }_{K_c}g^{\mbox{\tiny ind}}_{\mu\nu})+\partial_\mu X_b K^{\lambda}_c g^{\mbox{\tiny ind}}_{\lambda \nu}+\partial_\nu X_b K^{\lambda}_c g^{\mbox{\tiny ind}}_{\mu\lambda }\right)\;.\label{Afour}\eeqa
The first term  in (\ref{Afour}) vanishes since ${\cal L }_{K_c}g^{\mbox{\tiny ind}}_{\mu\nu}=0$, while the second term gives
\beqa
\epsilon^{abc}\partial_\mu X_b K^{\lambda}_c g^{\mbox{\tiny ind}}_{\lambda \nu} &=& \epsilon^{abc}\eta_{dr}\partial_\mu X_b K^{\lambda}_c \partial_\lambda X^d \partial_\nu X^r = \epsilon^{a}_{\ bc}\eta_{dr}\partial_\mu X^b \{X^c , X^d\} \partial_\nu X^r \nonumber \\
&=& \epsilon^{a}_{\ bc}\epsilon^{c}_{\ rs} X^s \partial_\mu X^b \partial_\nu X^r = - (\delta^a_{\ r}\eta_{bs} - \delta^a_{\ s}\eta_{br}) X^s \partial_\mu X^b \partial_\nu X^r \nonumber\\
&=&  - X_b \partial_\mu X^b \partial_\nu X^a + X^a g^{\mbox{\tiny ind}}_{\mu \nu} \equiv X^a g^{\mbox{\tiny ind}}_{\mu \nu} \;,
\eeqa
where in the last line we used (\ref{EXaXaismnsn}). Upon interchanging indicies $\mu$ and $\nu$, we see that the last term in (\ref{Afour}), $\epsilon^{abc}\partial_\nu X_b K^{\lambda}_c g^{\mbox{\tiny ind}}_{\mu\lambda }$, gives the same result.
We then arrive at  (\ref{rsclmetrc}).
}

\end{document}